\documentclass[11pt]{article}
\usepackage{amsmath}
\usepackage{amsfonts}
\usepackage{amssymb}
\usepackage{graphicx}

\def\bea{\begin{eqnarray}}
\def\eea{\end{eqnarray}}

\begin{document}
\begin{center}
\LARGE {\bf Cosmography in F(G) modified gravity }
\end{center}
\begin{center}
{\bf M. R. Setare\footnote{rezakord@ipm.ir} \\
  N. Mohammadipour\footnote{N.Mohammadipour@uok.ac.ir}}\\
 {\ Department of Science, University of Kurdistan \\
Sanandaj, IRAN.}

 \end{center}
\vskip 3cm
\begin{center}
{\bf{Abstract}}

Investigating the accelerated expansion of the universe with cosmography is a best method to constraint cosmological models. In this work, in the $F(G)$ modified gravity framework, we obtain equations of motion in a flat FRW metric. Then we reconstruct the present day values of $F(G)$ and its derivatives with the cosmographic parameters on the only assumption that the universe is homogenous and isotropic on large scale. Also we  investigate the conditions of cosmologically viable $F(G)$ gravity models with the fiducial data set values.
\end{center}

{\bf }

\newpage

\section{Introduction}

Recent accelerated expansion of our universe is one the most significant cosmological discoveries over the last decade \cite{1,2,3,4}. In order to explain this interesting behavior, many theories have been proposed. Although it is widely accepted that the cause which drives the acceleration is the so called dark energy, its nature and cosmological origin still remain enigmatic at present. The simplest suggestion for dark energy is cosmological constant. But it suffers from two kind of problems \cite{5} : fine tuning and coincidence problem. A dynamical scalar field with quintessence or phantom behavior is another proposal for dark energy ( for reviews see \cite{6} ).

Apart from the mentioned possibility of the existence of dark energy, a prominent possibility is that the gravitational interaction is modified at (at least) cosmic scales. Amongst the latter, models generalizing the Einstein-Hilbert action have been proposed. The Einstein field equations of General Relativity (GR) were first derived from an action principle by Hilbert, by adopting a linear functional of the scalar curvature, $R$, in the gravitational Lagrangian density. However, there are no a priori reasons to restrict the gravitational Lagrangian to this form, and indeed several generalizations of the Einstein-Hilbert Lagrangian have been proposed. The simplest way for modification of GR is to replace Ricci scalar, $R$ in Einstein- Hilbert action with a general function of the Ricci scalar which is well known as $F(R)$ gravity \cite{7,8}. For this kind of modification, one assumes that the gravitational action may contain some additional terms which starts to grow with decreasing curvature and obtain a late time acceleration epoch. One of the other of modification of Einstein's gravity is Gauss-Bonnet (GB) modification. As a possibility the Einstein-Gauss-Bonnet gravity is low energy limit of the string theory is of particular interest because of its special features. The GB generalization adds quadratic terms, involving second order curvature invariants (specifically Gauss-Bonnet term is a topological invariant in four dimensions) to the Einstein-Hilbert Lagrangian \cite{9}, (see also \cite{10}). As mentioned , in $4$-dimension the GB term is trivial. Therefore,
it is usually used from another form of modified gravity known as modified GB theory, e.g. see \cite{11}. In this theory, modification is done by adding an arbitrary function $F(G)$ into the Einstein-Hilbert Lagrangian. Also, in the present work, we focus on this model, but with an unknown function $F(G)$ in hand that must be determined for expanding accelerated universe in spatially flat FRW geometry. It is worth noting that both dark energy models and modified gravity theories have shown to be in agreement with the data.

However a possible solution could be to come back to the cosmography rather than finding out solutions of the Friedmann equations and testing them. In the present paper, we discuss the possibility to constrain $F(G)$ gravity models using a cosmographic approach \cite{12}. Cosmography relies  on two crucial things: I) extracting the maximum amount of information from measured distances, like the luminosity distances of SNeIa, II) assuming  that the universe is homogenous an isotropic on large scale. Recently, this approach was considered by using SN in \cite{13}, SN+GRBs in \cite{14} and SN+OHD+BAO in \cite{15}, where the current status of our universe can be read.

  This paper is outlined in the following manner: In Section \ref{2} we briefly review modified Gauss-Bonnet gravity, for self-completeness and self-consistency, and present the respective field equations. In section \ref{3} we introduce the basic notions of the cosmographic parameters then, write down all the quantities relevant of the $F(G)$ models (are presented in previous section) in terms of the cosmographic parameters. Inserting cosmographic parameters of $\Lambda$CDM model into the expressions of the $F(G)$ quantities, we will investigate in section \ref{4}. Section \ref{5} contains the main result of the paper demonstrating how the $F(G)$ derivatives can be related to the cosmographic parameters and we use these results and constraints on the cosmographic parameters to derive the cosmologically viable of $F(G)$ models and model independent estimates of the present day values of the $F(G)$ derivatives. Finally, in Section \ref{6}, we present our conclusions.

\section{Field equations and $F(G)$ modified gravity  }

We start with the 4-dimensional action in $F(G)$ gravity
\begin{equation}\label{1}
 S=\frac{1}{k^{2}}\int\sqrt{-g}\left[\frac{{\cal R}}{2}+F(G)\right]dx^4+ S_m,
\end{equation}
where ${\cal R}$ is Ricci scaler curvature, $F(G)$ is a generic function that must be determined\footnote{Indeed, $G$ is  topological invariant $G={\cal R}^2-4{\cal R}_{\alpha\beta}{\cal R}^{\alpha\beta}+{\cal R}_{\alpha\beta\gamma\delta}{\cal R}^{\alpha\beta\gamma\delta}$.}, $k^{2}=8\pi{\cal G}$ and $ S_m$ is the matter action.

Varying the action with respect to the metric, one can obtain the field equation as
\begin{eqnarray}\label{2}
G_{\mu \nu}
+8 [ {\cal R}_{\mu \rho \nu \sigma} +{\cal R}_{\rho \nu} g_{\sigma \mu}
+\frac{1}{2} (g_{\mu \nu} g_{\sigma \rho}
-g_{\mu \sigma} g_{\nu \rho}){\cal R} -{\cal R}_{\rho \sigma} g_{\nu \mu}
\\
\nonumber
 -{\cal R}_{\mu \nu} g_{\sigma \rho}+{\cal R}_{\mu \sigma} g_{\nu \rho} ] \nabla^{\rho} \nabla^{\sigma} F'(G)
+(G F'(G)-F) g_{\mu \nu}= T_{\mu \nu}^m,
\end{eqnarray}
where $G_{\mu\nu}$ the Einstein's tensor, $T_{\mu \nu}^m$ is the energy--momentum tensor of the matter, $k^{2}=8\pi{\cal G}=1$ and prime denotes the ordinary derivative with respect to $G$. For spatially flat Robertson-Walker metric
\begin{equation}\label{3}
{\rm d}s^2 = -{\rm d}t^2+a(t)^2 {\rm d}{\bf x}^2,
\end{equation}
we have
\begin{equation}\label{4}
{\cal R}=6(\dot H + 2H^2)\,,\quad\quad\quad G=24H^2 (\dot H + H^2)\,,
\end{equation}
where $H$ is the Hubble parameter and dots denotes the time
derivative. Considering that the universe is filled by pressureless dust matter, under the assumption of a flat universe the Eq.(\ref{4}) read
\begin{equation}\label{5}
3H^2 =G F'(G)-F-24H^3 \dot{F}'(G)+\rho_m,
\end{equation}
\begin{equation}\label{6}
-2\dot{H} =-8H^3 \dot{F}'(G)+16H\dot{H}\dot{F}'(G)+8H^2 \ddot{F}'(G)+\rho_m,
\end{equation}
and also continuity equations for the densities $\rho_{m}$ satisfy the following form
\begin{equation}\label{7}
\dot{\rho_m} + 3H\rho_m=0.
\end{equation}
Considering $F(G)$ gravity as  a factor that, is causing the accelerated expansion of the universe, we can rewrite the energy density and pressure of this dark energy as follows
\begin{equation}\label{8}
\rho_G =G F'(G)-F-24H^3 \dot{G} F''(G),
\end{equation}
\begin{eqnarray}\label{9}
p_G =[(16H^3 \dot{G}+16H\dot{H}\dot{G}+8H^2 \ddot{G})F''(G)\\
\nonumber
+8H^2\dot{G}^2F'''(G)-G F'(G)+F].~~~~~
\end{eqnarray}
 The continuity equation for $\rho_{G}$ leads to
\begin{equation}\label{10}
\dot{\rho}_G+3H(p_G+\rho_G)=0;
\end{equation}
with the equation of state as
\begin{equation}\label{11}
w_{G}=-1+\frac{(8H^2\ddot{G}+16H\dot{H}\dot{G}-8H^3\dot{G})F''+8H^{2}\dot{G}^{2}F'''}{GF'-F-24H^3\dot{G}F''}.
\end{equation}
Here, quantities $\rho_{G}$ and $\omega_{G}$ depends on $H$, $G$, $F(G)$ and there derivatives. These quantities could be related to the observed acceleration of the universe.
\section{F(G) gravity vs cosmography }

In this section, firstly defined cosmographic parameters then, instead of solving Eq.(\ref{6}) for given values of the boundary conditions, we obtain the relation between the present day values of $F(G)$, its derivatives and cosmographic parameters.
\begin{center}
{\bf{A.Cosmography parameters}}
\end{center}

The cosmographic parameters that are proportional to the coefficients Taylor series expansion of the scale factor with respect to the cosmic time defined as
\begin{eqnarray}\label{12}
H(t)&=&\frac{1}{a}\frac{da}{dt},\\
q(t)&=&-\frac{1}{H^{2}}\frac{1}{a}\frac{d^{2}a}{dt^{2}},\\
j(t)&=&\frac{1}{H^{3}}\frac{1}{a}\frac{d^{3}a}{dt^{3}},\\
s(t)&=&\frac{1}{H^{4}}\frac{1}{a}\frac{d^{4}a}{dt^{4}},\\
l(t)&=&\frac{1}{H^{5}}\frac{1}{a}\frac{d^{5}a}{dt^{5}},
\end{eqnarray}
which are usually referred to as the Hubble, deceleration, jerk, snap and lerk parameters, respectively. For example, the present value
of Hubble parameter $H_{0}$ describes the present expansion rate of our universe, and a negative value of $q_{0}$ means that our
universe is undergoing an accelerated expansion while, $j_{0}$ allows to discriminate among
different accelerating models. We can rewrite the Hubble parameter and its derivatives in terms of cosmographic parameters as
\begin{eqnarray}\label{17}
\dot{H}=-H^{2}(1+q),
\end{eqnarray}
\begin{eqnarray}\label{18}
\ddot{H}=H^{3}(j+3q+2),
\end{eqnarray}
\begin{eqnarray}\label{19}
\frac{d^{3}H}{dt^{3}}=H^{4}[s-4j-3q(q+4)-6,
\end{eqnarray}
\begin{eqnarray}\label{20}
\frac{d^{4}H}{dt^{4}}=H^{5}[l-5s+10(q+2)j+30(q+2)q+24],
\end{eqnarray}
where a dot denotes derivative with respect to the cosmic time. Using Eqs.(\ref{17})-(\ref{20}) we can rewrite the Gauss-Bonnet term and its derivatives in terms of cosmographic parameters.
\begin{center}
{\bf{B.F(G)-derivatives and cosmography}}
\end{center}

Rather than solving Eq.(\ref{6}) and choosing $F(G)$ by given values of boundary conditions, we reconstruct the present day value of $F(G)$ and its derivatives with the present time of the cosmographic parameters $(H_{0}, q_{0}, j_{o,}, s_{o}, l_{o})$.

Considering Eq.(\ref{4}) and differentiating with respect to $t$, one can obtain the following equations
\begin{eqnarray}\label{21}
\dot{G}=24[2(2H^{3}+H\dot{H})\dot{H}+H^{2}\ddot{H}],
\end{eqnarray}
\begin{eqnarray}\label{22}
\ddot{G}=24[2(6H^{2}+\dot{H})\dot{H}^{2}+2(2H^{3}+3H\dot{H})\ddot{H}+H^{2}\frac{d^{3}H}{dt^{3}}],
\end{eqnarray}
\begin{eqnarray}\label{23}
\frac{d^{3}G}{dt3}=24[24H\dot{H}^{3}+6(6H^{2}\dot{H}+2\dot{H}^{2}+H\ddot{H})\ddot{H}\\
\nonumber
+4(H^{3}+2H\dot{H})\frac{d^{3}H}{dt^3}+H^2\frac{d^4 H}{dt^4}].~~~~~~~~
\end{eqnarray}
Since we will reconstruct the present day value of $F(G)$ and its derivatives, then rewrite Eqs.(\ref{5}), (\ref{6}) at the present time $t=t_{0}$ as
\begin{eqnarray}\label{24}
H_{0}^{2} =\frac{G_{0} F'(G_{0})}{3}-\frac{F(G_{0})}{3}-8H_{0}^3 \dot{G_{0}}F''(G_{0})+H_{0}^{2}\Omega_{0m},
\end{eqnarray}
\begin{eqnarray}\label{25}
\dot{H_{0}}=(4H_{0}^3 \dot{G_{0}}-8H_{0}\dot{H_{0}}\dot{G_{0}}-4H_{0}^2 \ddot{G_{0}})F''(G_{0})\\
\nonumber
-4H_{0}^2\dot{G_{0}}^2F'''(G_{0})-\frac{3H_{0}^{2}}{2}\Omega_{0m},~~~~~~~~~~~~
\end{eqnarray}
where $\rho_{m}=\rho_{m}(t=t_{0})a^{-3}=3H_{0}^{2}\Omega_{0m}a^{-3}$, with $\Omega_{0m}$ the present day matter density which is obtained from the matter continuity equation.

We will assume that $F(G)$ may be well approximated by its third order Taylor expansion in $(G-G_{0})$ following as
\begin{eqnarray}\label{26}
F(G)=F(G_{0})+F'(G_{0})(G-G_{0})+\frac{1}{2}F''(G_{0})(G-G_{0})^{2}\\
\nonumber
+\frac{1}{6}F'''(G_{0})(G-G_{0})^{3},~~~~~~~~~~~~~~~~~~~~~~~~~~~~~~~~~
\end{eqnarray}
with $G_{0}$ being the current value of $G$. Here we assume $\frac{d^{n}F(G)}{dG^{n}}=0$ for $n\geq4$. In order to get the third derivative of $F(G)$, we differentiate of Eq.(\ref{6}) with respect to $t$ and evaluate in $(t=t_{0})$. One finally obtain
\begin{eqnarray}\label{27}
\ddot{H_{0}}=\frac{9H_{0}^3}{2}\Omega_{0m}+4[(3H_{0}^2\dot{H_{0}}-2\dot{H_{0}}^2-2H_{0}\ddot{H_{0}})\dot{G_{0}}+~~~~~~~~~~~~\\
\nonumber
(H_{0}^3-4H_{0}\dot{H_{0}})\ddot{G_{0}}-H_{0}^2\frac{d^3 G_{0}}{dt^3}]F''(G_{0})-~~~~~~~~~~~~~~~~~~~\\
\nonumber
4[3H_{0}^2\dot{G_{0}}\ddot{G_{0}}-(H_{0}^3-4H_{0}\dot{H_{0}})\dot{G_{0}}^{2}]F'''(G_{0}).~~~~~~~~~~~~~~~
\end{eqnarray}
Evaluate the present day value of ($\dot{H}$, $\ddot{H}$, $\frac{d^{3}H}{dt^{3}}$, $\frac{d^{4}H}{dt^{4}}$) Eqs.(\ref{17})-(\ref{20}) and substitute these equations into Eq.(\ref{4}) and Eqs.(\ref{21})-(\ref{23}) to obtain $G$ and its derivatives at the present time:
\begin{eqnarray}\label{28}
G_{0}=-24H_{0}^4q_{0},
\end{eqnarray}
\begin{eqnarray}\label{29}
\dot{G_{0}}=24H_{0}^5(2q_{0}^2+3q_{0}+j_{0}),
\end{eqnarray}
\begin{eqnarray}\label{30}
\ddot{G_{0}}=24H_{0}^6[(q_{0}+1)(-2q_{0}^2-12q_{0}-6j_{0})-q_{0}^2+s_{0}],
\end{eqnarray}
\begin{eqnarray}\label{31}
\frac{d^3 G_{0}}{dt^3}=24H_{0}^6[(q_{0}+1)(36q_{0}^2+60q_{0}-8s_{0})+12q_{0}^2\\
\nonumber
+l_{0}-s_{0}+6j_{0}(2q_{0}^2+11q_{0}+j_{0}+6)].~~~~~
\end{eqnarray}
Using Eqs.(\ref{17})-(\ref{20}) and by inserting Eqs.(\ref{28})-(\ref{31}) into Eq.(\ref{25}) and Eq.(\ref{27}) finally, we reconstruct the present day value of the second and the third order derivatives of $F(G)$ with respect to cosmographic parameters at $z=0$ as
\begin{eqnarray}\label{32}
\frac{F''(G_{0})}{(24H_{0}^{6})^{-1}}=\frac{A_{02}+B_{02}\Omega_{0m}}{C_{0}},
\end{eqnarray}
\begin{eqnarray}\label{33}
\frac{F'''(G_{0})}{(24H_{0}^{5})^{-2}}=\frac{A_{03}+B_{03}\Omega_{0m}}{C_{0}},
\end{eqnarray}
with
\begin{eqnarray}\label{34}
A_{02}=4(2q_{0}^{2}+3q_{0}+j_{0})(j_{0}^{2}-21j_{0}-39q_{0}j_{0}-20q^{2}_{0}j_{0}\\
\nonumber
-75q_{0}^{3}+3q_{0}s_{0}-14q_{0}^{4}-105q_{0}^{2}+3s_{0}-45q_{0}),~
\end{eqnarray}
\begin{eqnarray}\label{35}
B_{02}=-6(2q_{0}^{2}+3q_{0}+j_{0})(16j_{0}+14q_{0}j_{0}+29q_{0}^{2}-3s_{0}-2q_{0}^{3}+30q_{0}),
\end{eqnarray}
\begin{eqnarray}\label{36}
A_{03}=48s_{0}-228j_{0}-4l_{0}-420q_{0}-888q_{0}^{2}-408q_{0}j_{0}-762q_{0}^{3}-298q_{0}^{2}j_{0}\\
\nonumber
+4j_{0}^{2}-404q_{0}^{4}-144q_{0}^{3}j_{0}-48q_{0}^{5}+48q_{0}^{2}s_{0}-4j_{0}s_{0}+60q_{0}s_{0}-4q_{0}l_{0},
\end{eqnarray}
\begin{eqnarray}\label{37}
B_{03}=288j_{0}+6l_{0}-66s_{0}+504q_{0}+1242q_{0}^{2}+690q_{0}j_{0}\\
\nonumber
+720q_{0}^{3}+252q_{0}^{2}j_{0}+48j_{0}^{2}-72q_{0}s_{0}+72q_{0}^{4},~~~~~
\end{eqnarray}
\begin{eqnarray}\label{38}
C_{0}=-16(2q_{0}^{2}+3q_{0}+j_{0})(8j_{0}^{3}-132j_{0}^{2}-219j_{0}^{2}q_{0}-118q_{0}^{2}j_{0}^{2}\\
\nonumber
+36j_{0}s_{0}+j_{0}l_{0}-570q_{0}j_{0}-148q_{0}^{4}j_{0}-1239q_{0}^{2}j_{0}-3s_{0}^{2}\\
\nonumber
-852q_{0}^{3}j_{0}+34j_{0}q_{0}s_{0}-1626q_{0}^{3}-630q_{0}^{2}+84q_{0}^{2}s_{0}+3q_{0}l_{0}\\
\nonumber
+72q_{0}s_{0}-1419q_{0}^{4}-60q_{0}^{6}+2q_{0}^{2}l_{0}-468q_{0}^{5}+8q_{0}^{3}s_{0}).~~~
\end{eqnarray}
Insert Eq.(\ref{32}) into Eq.(\ref{24}) and we get to relation between the present day value of $F(G)$ and its first derivative
\begin{eqnarray}\label{39}
\frac{F'(G_{0})}{(24H_{0}^{2}q_{0}^{4})^{-1}}+\frac{F(G_{0})}{H_{0}^{2}}=-3\frac{(C_{0}+8a_{0}A_{02})+(8a_{0}B_{02}-C_{0})\Omega_{0m}}{C_{0}},
\end{eqnarray}
where
\begin{eqnarray}\label{40}
a_{0}=(2q_{0}^{2}+3q_{0}+j_{0}).
\end{eqnarray}

Eqs.(\ref{32})-(\ref{40}) make it possible to estimate the present day values of the second, the third derivatives of $F(G)$ and the relation between $F(G)$ and its first derivative as function of the Hubble constant $H_{0}$ and the cosmographic parameters ($q_{0}, j_{0}, s_{0}, l_{0}$) provided a value for the matter density parameter $\Omega_{om}$ is given. Although, the value of $\Omega_{0m}$ is usually the result of fitting a given dataset in the framework of an assumed dark energy scenario, but different models all converge towards the concordance value $\Omega_{om}\simeq 0.25$ which is also in agreement with astrophysical (model independent) estimates from the gas mass fraction in galaxy clusters. Indeed, Eqs.(\ref{32})-(\ref{40}) are ideal tools to testing $F(G)$-gravity models with observational dataset .

\section{The $\Lambda$CDM model}

We can test the reliability by comparing result are obtained in previous section with spatially flat $\Lambda$CDM model. The cosmographic parameters for the $\Lambda$CDM model with equation of state parameter $\omega=-1$ at the present time read
\begin{eqnarray}\label{41}
q_{0}=-1+\frac{3}{2}\Omega_{0m},
\end{eqnarray}
\begin{eqnarray}\label{42}
j_{0}=1,
\end{eqnarray}
\begin{eqnarray}\label{43}
s_{0}=1-\frac{9}{2}\Omega_{0m},
\end{eqnarray}
\begin{eqnarray}\label{44}
l_{0}=1+3\Omega_{0m}+\frac{27}{2}\Omega_{0m}^{2}.
\end{eqnarray}
Inserting Eqs.(\ref{41})-(\ref{44}) into Eqs.(\ref{32}) and (\ref{33}) gives
\begin{eqnarray}\label{45}
F''(G_{0})=F'''(G_{0})=0.
\end{eqnarray}
From Eq.(\ref{31}) and the previous equations in Eqs.(\ref{41})-(\ref{44}) one can obtain
\begin{eqnarray}\label{46}
F(G_{0})=-\Lambda+G_{0}F'(G_{0}).
\end{eqnarray}

The equation of state for these models defined by Eq.(\ref{11}) that, by using Eq.(\ref{45}) reduce to $\omega_{G}=-1$. Thus, we obviously conclude that the $R+F(G)$ theory having exactly the same cosmographic parameters as the $\Lambda$CDM model with equation of state $\omega_{G}=-1$ if $G_{0}F'(G_{0})-F(G_{0})=\Lambda$. If we assume $G_{0}F'(G_{0})=2\Lambda$ then $F(G_{0})= \Lambda$ and the action Eq.(\ref{1}) reduce to GR\,+\,$\Lambda$\,+\,matter.\\
\section{Constraining $F(G)$ parameters}

In order to constrain the model by cosmography, i.e. to estimate the function $G_{0}F'(G_{0})-F(G_{0})$, $F''(G_{0})$ and $F'''(G_{0})$ at the present time, we need to constrain observationally the cosmographic parameters by using appropriate distance indicators. Moreover, we must take
care that the expansion of the distance related quantities in terms of $(q_{0}, j_{0}, s_{0}, l_{0})$ closely follows the exact expressions over the range probed by the data used.
\begin{center}
\bf {A. Observational constraints }
\end{center}

Taking SNeIa and a fiducial $\Lambda$CDM model as a test case, one has to check that the approximated luminosity distance\footnote{See \cite{12} for the analytical expression.} deviates from the $\Lambda$CDM one less than the measurement uncertainties up to $z \simeq  1.5$ to avoid introducing any systematic bias. Since we are interested in constraining $(q_0, j_0, s_0, l_0)$, we will expand the luminosity distance $D_L$ up to the fifth order in $z$ which indeed allows us to track the $\Lambda$CDM expression with an error less than $1\%$ over the full redshift range. We have checked that this is the case also for the angular diameter distance $D_A = D_L(z)/(1 + z)^2$ and the Hubble parameter $H(z)$ which, however, we expand only up to the fourth order to avoid introducing a further cosmographic parameter. In order to constrain the parameters $(h, q_0, j_0, s_0, l_0)$, we consider the dataset that is used in \cite{12} and are summarized in Table\,\ref{tab: 1}.

\begin{table}[t]
\begin{center}
\begin{tabular}{cccccc}
\hline
$x$ & $x_{BF}$ & $\langle x \rangle$ & $x_{med}$ & $68\%$ CL & $95\%$ CL \\
\hline \hline
~ & ~ & ~ & ~ & ~ & ~ \\
$h$ & 0.718& 0.706 & 0.706 & (0.693, 0.719) & (0.679, 0.731) \\
~ & ~ & ~ & ~ & ~ & ~ \\
$q_0$ & -0.64 & -0.44 & -0.43 & (-0.60, -0.30) & (-0.71, -0.26) \\
~ & ~ & ~ & ~ & ~ & ~ \\
$j_0$ & 1.02 & -0.04 & -0.15 & (-0.88, -0.90) & (-1.07,  1.40) \\
~ & ~ & ~ & ~ & ~ & ~ \\
$s_0$ & -0.39 & 0.18 & 0.02 & (-0.57,  1.07) & (-1.04,  1.78) \\
~ & ~ & ~ & ~ & ~ & ~ \\
$l_0$ & 4.05 & 4.64 & 4.54 & (2.99,  6.48) & (1.78,   8.69) \\
~ & ~ & ~ & ~ & ~ & ~ \\
\hline
\end{tabular}
\end{center}
\caption{Constraints on the cosmographic parameters by jointly fitting the Union2 SNeIa sample and the BAO data. Columns are as follows\,: 1. parameter id; 2. best fit; 3., 4. mean and
median from the marginalized likelihood; 5., 6. $68$ and $95\%$ confidence ranges.}
\label{tab: 1}
\end{table}

In order to translate our constraints on the cosmographic parameters on similar constraints on $F(G)$ and its derivatives, we should just use Eqs.(\ref{32}), (\ref{33}) and (\ref{39}) evaluating them along the final coadded and thinned chain and then looking at the corresponding histograms. We define quantities for shortness as
\begin{eqnarray}\label{47}
f_{0}=G_{0}F'(G_{0})-F(G_{0}),\hspace{1cm}f_{2}=F''(G_{0}), \hspace{1cm} f_{3}=F'''(G_{0}).
\end{eqnarray}

We constrain these quantities by setting the value of $\Omega_{0m}$ along the chain using $\Omega_{0m}=\omega_{m}h^{-2}$ with the physical matter density $\omega_{m}=0.1329$ in agreement with the WMAP7 data. For each value of $h$ along the chain, we fix $\Omega_{m0} = \omega_m/h^2$ having neglected the error on $\omega_m$ since it is subdominant with respect to the one on $h$. We stress that, although the fiducial value for $\omega_m$ has been obtained for a $\Lambda$CDM model, it should be unchanged for any model which reduces to the GR\,+\,matter domination at the CMBR epoch as is our case.

We finally analyize the present day value of quantities summarized in Table\,\ref{tab: 2}. It is worth noting that the $f_{i}$ values become smaller as the order n of the derivative increases and it is in agreement with our assumption in Eq.(\ref{26}). Note also that, mean and median from the marginalized likelihood values of $(f_{2}, f_{3})$ are almost equal and are not quite different from their median values. As a further remark, we note that the confidence ranges of constraints on $f_{0}, f_{2}, f_{3}$ (in particular, for $95\%$) become larger as the order $i$ of the derivative increases. Which caused the degeneracies among $(q_0, j_0, s_0, l_0)$ and the nonlinear behavior of the relations $f_{i}$.

\begin{table}[t]
\begin{center}
\begin{tabular}{cccccc}
\hline
$x$ & $x_{BF}$ & $\langle x \rangle$ & $x_{med}$ & $68\%$ CL & $95\%$ CL \\
\hline \hline
~ & ~ & ~ & ~ & ~ & ~ \\
$f_{0}$ & 1.139 & 0.921 & 0.905 & (0.859, 0.854) & (0.765, 1.801) \\
~ & ~ & ~ & ~ & ~ & ~ \\
$f_{2}$ & 0.0027 & 0.0050 & 0.0050 & (0.0030, 0.0045) & (0.0038, 0.0321) \\
~ & ~ & ~ & ~ & ~ & ~ \\
$f_{3}$ & -0.0231 & -0.0020 & -0.0019 & (-0.0011, -0.0010) & (-0.0015,  0.0734) \\
~ & ~ & ~ & ~ & ~ & ~ \\
\hline
\hline
~ & ~ & ~ & ~ & ~ & ~ \\
$w_{G_{0}}$ & -1.0238 & -0.8542 & -0.8454 & (-1.0139, -0.0718) & (-1.1334, -0.9849 ) \\
~ & ~ & ~ & ~ & ~ & ~ \\
\hline
\end{tabular}
\end{center}
\caption{Constraints on quantities $f_i$ and equation of state values from the Markov Chain for the cosmographic parameters. Columns order is the same as in Table\,\ref{tab: 1}.}
\label{tab: 2}
\end{table}

\begin{center}
\bf{B. Observational constraints on cosmologically viable $F(G)$ gravity models}
\end{center}

In this section, we translate the constraints on the cosmographic parameters on similar constraints on the present day values of conditions cosmologically viable $F(G)$ models. De Felice et al. \cite{16} have investigated the viability of cosmological $F(G)$ gravity models. They started work with our action Eq.(\ref{1}) in a spatially FRW background with the metric signature (-, +, +, +) and showed that the viable of this models need to satisfy the following conditions:
\begin{itemize}
\item (1) $F(G)$ and its derivatives $F'(G)$, $F''(G)$,...
are regular.
\item (2) $F''(G)>0$ for $G \le G_{1}$ and $F''(G)$
approaches $+0$ in the limit $|G| \to \infty$.
\item (3) $0<H_1^6 F''(G_{1})<1/384 \simeq 0.0026$ is the stable condition at the de Sitter point.
\item (4) $0<H_1^6 F''(G_{1})<1/600 \simeq 0.0017$ corresponds to a stable spiral (damping with oscillations).
\end{itemize}

Here index $1$, in this context, denotes the calculated values of this quantities in de-Sitter point. In order to investigate the above conditions with the cosmographic parameters at the present time, we use Table \ref{tab: 1} and evaluate $M_{0}=H_{0}^6F''(G_{0})$. Finally, the present day values of $f_{2}=F''(G_{0})$ and $M_{0}$ summarize in Table\ref{tab: 3}.

The observational constraints summarized in Table\ref{tab: 3} shows that, the present day value of $f_{2}=F''(G_{0})>0$ and it is in agreement with the similar results in \cite{16} to have a stable de \,-\ Sitter point in $F(G)$ models. We also note that the values $M_{0}=H_{0}^6F''(G_{0})$ in Table\ref{tab: 3} corresponds to a stable spiral de \,-\ Sitter point (damping with oscillations).
\begin{table}[t]
\begin{center}
\begin{tabular}{cccccc}
\hline
$x$ & $x_{BF}$ & $\langle x \rangle$ & $x_{med}$ & $68\%$ CL & $95\%$ CL \\
\hline \hline
~ & ~ & ~ & ~ & ~ & ~ \\
$f_{2}$ & 0.0027 & 0.0050 & 0.0050 & (0.0030, -0.0050) & (0.0038, 0.0321) \\
~ & ~ & ~ & ~ & ~ & ~ \\
$H^6_{0}f_{2}$ & 0.0004 & 0.0006 & 0.0006 & (0.0003, -0.0007) & (0.0004,  0.005) \\
~ & ~ & ~ & ~ & ~ & ~ \\
\hline
\end{tabular}
\end{center}
\caption{Constraints on $M_0$ values from the Markov Chain for the cosmographic parameters. Columns order is the same as in Table\,\ref{tab: 1}.}
\label{tab: 3}
\end{table}

\begin{center}
\bf{C. Cosmography VS $F(G)$ models}
\end{center}

Up to now, for a generic function of $F(G)$, we obtain observational constraints on $F(G)$ and its derivatives from cosmographic parameters. Now, we check the viability of a given $F(G)$ model with constraints are provided in previous section without the need of explicitly solving the field equations and fitting data.

As an example, we assume the following model as\cite{17}
\begin{eqnarray}\label{48}
F(G)=\alpha G^{n}+\beta G\ln G,
\end{eqnarray}
where $\alpha$, $\beta$ and $n$ are constants. Using quantities Eq.(\ref{47}) and Eqs.(\ref{32}), (\ref{33}) and (\ref{39}), one can obtain the following expressions
\begin{eqnarray}\label{49}
f_{0}&=&\alpha(n-1)G_{0}^{n}+\beta G_{0},\\
f_{2}&=&\alpha n(n-1)G_{0}^{n-2}+ \beta G_{0}^{-1},\\
f_{3}&=&\alpha n(n-1)(n-2)G_{0}^{n-3}-\beta G_{0}^{-2}.
\end{eqnarray}
Solving the first and second equations with respect to $(\alpha, \beta)$ leads to
\begin{eqnarray}\label{50}
\alpha&=&-\frac{f_{0}-G_{0}^{2}f_{2}}{(n-1)^{2}}G_{0}^{-n},\\
\beta&=&\frac{nf_{0}-G_{0}^{2}f_{2}}{n-1}G_{0}^{-1}.
\end{eqnarray}
Inserting $\alpha$ and $\beta$ into the third equation of Eq.(51), we obtain two set solutions for $n$.
\begin{eqnarray}\label{51}
n=\frac{P(f_{0}, f_{2}, f_{3}, G_{0})\pm\sqrt{Q(f_{0}, f_{2}, f_{3}, G_{0})}}{T(f_{0}, f_{2}, G_{0})},
\end{eqnarray}
with
\begin{eqnarray}\label{52}
P(f_{0}, f_{2}, f_{3}, G_{0})&=&f_{3}+2G_{0}^{-1}f_{2}-G_{0}^{-3}f_{0},\\~~~~~~~~
T(f_{0}, f_{2}, G_{0})&=&f_{2}G_{0}^{-1}-f_{0}G_{0}^{-3},~~~~~~~~
\end{eqnarray}
\begin{eqnarray}\label{53}
Q(f_{0}, f_{2}, f_{3}, G_{0})=G_{0}^{-6}f_{0}^{2}+4G_{0}^{-2}f_{2}^{2}+f_{3}^{2}-4f_{0}f_{2}G_{0}^{-4}\\
\nonumber
+2f_{0}f_{3}G_{0}^{-3}-4f_{2}f_{3}G_{0}^{-1}.~~~~~~~~~~~~~
\end{eqnarray}
We can see that values of $(\alpha, \beta, n)$ depends on the values of $(q_0, j_0, s_0, l_0)$ then, one may obtain one, two or any acceptable solution, i.e. real positive values of $n$. If the final values of $(\alpha, \beta, n)$ are physically viable, we can conclude that the model in Eq.(\ref{48}) is in agreement with the data giving the same cosmographic parameters inferred from the data themselves.
\section{ Conclusion }

Cosmography is a ideal way to give a picture of the observed universe considering only assumptions that, the universe is homogenous and isotropic on large scale. In the present paper we have discussed
the cosmography of a modified GB model with an unknown function of topological invariant GB. At first we have defined cosmographic parameters, then instead of solving equations of motion for given values of the boundary conditions, we have obtained the relation between the present day values of $F(G)$, its derivatives and cosmographic parameters. We have assumed that $F(G)$ may be well approximated by its third order Taylor expansion in $(G-G_{0})$. Then  we have reconstructed the present day value of the second and the third order derivatives of $F(G)$ with respect to cosmographic parameters at $z=0$ by Eqs.(\ref{32}), (\ref{33}). We also found a relation between the present day value of $F(G)$ and its first derivative. This recent relation and also the second and the third order derivatives of $F(G)$  are functions of the Hubble constant $H_{0}$ and the cosmographic parameters ($q_{0}, j_{0}, s_{0}, l_{0}$), and $\Omega_{om}$ .\\
In contrast to the cosmography investigations in $f(R)$ and $f(T)$ \cite{12} here we note that, in our model do not have to considering this assumption that $G_{eff}(z=0)=G_N\rightarrow f'(0)=1$.
We transfer the constraints on $(H_{0}, q_0, j_0, s_0, l_0)$ and $\Omega_{om}$ into similar ones for $f_{i}$ quantities and summarized them in Table\,\ref{tab: 2}. It is worth noting that, these constraints (comes out from $f_{i}$) showed that, the validity of our assumption in the Taylor expansion of $F(G)$ which is $\frac{d^{n}F(G)}{dG^{n}}=0$ for $n\geq4$. On the other hand, it is clear that  the increasing  order of expansion shifts away from the $\Lambda$CDM fiducial values. Then, it is possible recover the $\Lambda$CDM with the above assumption in $F(G)$ gravity.

As a further remark, we have investigated the conditions of cosmologically viable $F(G)$ gravity models with the fiducial values in Table\,\ref{tab: 3}. These values were in agreement with the viable conditions of this model that the authors of \cite{16} investigated, so that the stable condition is $f_{2}=F''(G)>0$ and $0<H_1^6 F''(G_{1})<1/600 \simeq 0.0017$ corresponds to a stable spiral (damping with oscillations).


\begin{thebibliography}{99}
\bibitem{1}S. Perlmutter et al. [Supernova Cosmology Project Collaboration], Astrophys. J. 517,
565 (1999).
\bibitem{2}C. L. Bennett et al., Astrophys. J. Suppl. 148, 1 (2003).
\bibitem{3}M. Tegmark et al. [SDSS Collaboration], Phys. Rev. D 69, 103501 (2004).
\bibitem{4}S. W. Allen, et al., Mon. Not. Roy. Astron. Soc. 353, 457 (2004).
\bibitem{5}S. Weinberg, Rev. Mod. Phys. 61, 1 (1989); V. Sahni and A. Starobinsky,
Int. J. Mod. Phys. D 9, 373 (2000); P. J. Peebles and B. Ratra,
Rev. Mod. Phys. 75, 559 (2003).
\bibitem{6}E. J. Copeland, M. Sami and S. Tsujikawa Int. J. Mod. Phys. D 15,
1753 (2006); Y. F. Cai, E. N. Saridakis, M. R. Setare, J. Q. Xia, Phys.
Rep. 493, 1, 1-60 (2010).
\bibitem{7}S. Nojiri and S. D. Odintsov, Int. J. Geom. Meth. Mod. Phys. 4, 115 (2007).
\bibitem{8}S. Nojiri and S. D. Odintsov, arXiv:0801.4843 [astro-ph]; arXiv:0807.0685 [hep-th]; T.
P. Sotiriou and V. Faraoni, arXiv:0805.1726 [gr-qc]; F. S. N. Lobo, arXiv:0807.1640 [grqc];
S. Capozziello and M. Francaviglia, Gen. Rel. Grav. 40, 357 (2008); M. R. Setare,
Int. J. Mod. Phys. D17, 2219, (2008).
\bibitem{9}B. M. N. Carter, I. P. Neupane, Phys. Lett. B 638, 94 (2006).
\bibitem{10}S. M. Carroll, A. De Felice, V. Duvvuri, D. A. Easson, M. Trodden and M. S. Turner,
Phys. Rev. D 71, 063513 (2005).
\bibitem{11}S. Nojiri and S. D. Odintsov, Phys. Lett. B 631, 1 (2005); S. Nojiri,
S. D. Odintsov and O. G. Gorbunova J. Phys. A: Math. Gen. 39, 6627
(2006); G. Cognola, E. Elizalde, S. Nojiri, S. D. Odintsov and S. Zerbini
Phys. Rev. D 73, 084007 (2006).
\bibitem{12}S. Capozziello, V. F. Cardone and V. Salzano, Phys. Rev. D 78 (2008) 063504 [arXiv:0802.1583 [astro-ph]]; S. Capozziello, V. F. Cardone, [arXiv:0902.0088 [astro-ph]]; M. Bouhmadi\,-\,Lopez, S. Capozziello, V. F. Cardone, [arXiv:1010.1547 [gr-qc]]; S. Capozziello1, V. F. Cardone, H. Farajollahi, A. Ravanpak, [arXiv:1108.2789 [astro-ph]]
\bibitem{13}A. C. C. Guimaraes, J. V. Cunha and J. A. S. Lima, [arXiv:0904.3550].
\bibitem{14}V. Vitagliano, J. Q. Xia, S. Liberati, M. Viel, JCAP 03 (2010) 005 [arXiv:0911.1249v2 [astro-ph.CO]].
\bibitem{15}S. Capozziello, R. Lazkoz, V. Salzano, [arXiv :1104.3096, 2011]; L. Xu, W. Li, J. Lu, JCAP 0907,031(2009) [arXiv:0905.4552v1 [astro-ph.CO]].
\bibitem{16}A. De Felice, S. Tsujikawa, Phys. Lett. B 675 (2009) 1-8, [arXiv:0810.5712v2 [hep-th]].
\bibitem{17}H.-J. Schmidt, Phys. Rev. D 83 (2011) 083513, [	arXiv:1102.0241v2 [gr-qc]].
\end{thebibliography}
\end{document}